\documentclass[aps,prl,
showpacs,
superscriptaddress,reprint,floatfix,citeautoscript,twocolumn]{revtex4-2}
\usepackage{graphicx}
\newcommand{\RNum}[1]{\uppercase\expandafter{\romannumeral #1\relax}}
\usepackage{amsmath}
\usepackage{amssymb}
\usepackage{bm}
\usepackage{float}
\usepackage{dcolumn}
\usepackage{subfigure}
\usepackage{units}
\usepackage{hyperref}
\usepackage[usenames]{color}
\usepackage[dvipsnames]{xcolor}
\usepackage{booktabs}
\usepackage{multirow}
\usepackage{siunitx}
\usepackage{wrapfig}
\usepackage{lipsum}
\usepackage{setspace}
\usepackage{xcolor}
\usepackage{longtable}
\hypersetup{pdfborder=0 0 0,colorlinks=true,citecolor=blue,linkcolor=blue}
\input epsf

\begin{document}

\title{Spin wave dispersion of ultra-low damping hematite ($\alpha\text{-Fe}_2\text{O}_3$) at GHz frequencies}
\author{Mohammad Hamdi}
\email[]{mohammad.hamdi@epfl.ch, \\
	mohamad.hamdi90@gmail.com}
\affiliation{\'Ecole Polytechnique F\'ed\'erale de Lausanne (EPFL),  Institute of Materials, Laboratory of Nanoscale Magnetic Materials and Magnonics,  CH-1015 Lausanne, Switzerland}

\author{Ferdinand Posva}
\affiliation{\'Ecole Polytechnique F\'ed\'erale de Lausanne (EPFL),  Institute of Materials, Laboratory of Nanoscale Magnetic Materials and Magnonics,  CH-1015 Lausanne, Switzerland}

\author{Dirk Grundler}
\email[]{dirk.grundler@epfl.ch}
\affiliation{\'Ecole Polytechnique F\'ed\'erale de Lausanne (EPFL),  Institute of Materials, Laboratory of Nanoscale Magnetic Materials and Magnonics,  CH-1015 Lausanne, Switzerland}
\affiliation{\'Ecole Polytechnique F\'ed\'erale de Lausanne (EPFL), Institute of Electrical and Micro Engineering, CH-1015 Lausanne, Switzerland}

\date{\today}

\begin{abstract}
Low magnetic damping and high group velocity of spin waves (SWs) or magnons are two crucial parameters for functional magnonic devices. Magnonics research on signal processing and wave-based computation at GHz frequencies focussed on the artificial ferrimagnetic garnet Y$_3$Fe$_5$O$_{12}$ (YIG) so far. We report on spin-wave spectroscopy studies performed on the natural mineral hematite ($\alpha\text{-Fe}_2\text{O}_3$) which is a canted antiferromagnet. By means of broadband GHz spectroscopy and inelastic light scattering, we determine a damping coefficient of $1.1\times10^{-5}$ and magnon group velocities of a few 10 km/s, respectively, at room temperature. Covering a large regime of wave vectors up to $k\approx 24~{\rm rad}/\mu$m, we find the exchange stiffness length to be relatively short and only about 1 \AA. In a small magnetic field of 30 mT, the decay length of SWs is estimated to be 1.1 cm similar to the best YIG. Still, inelastic light scattering provides surprisingly broad and partly asymmetric resonance peaks. Their characteristic shape is induced by the large group velocities, low damping and distribution of incident angles inside the laser beam. Our results promote hematite as an alternative and sustainable basis for magnonic devices with fast speeds and low losses based on a stable natural mineral.


\end{abstract}

\maketitle

\begin{figure}[!h]
	\centering
	\includegraphics[width=0.4\textwidth]{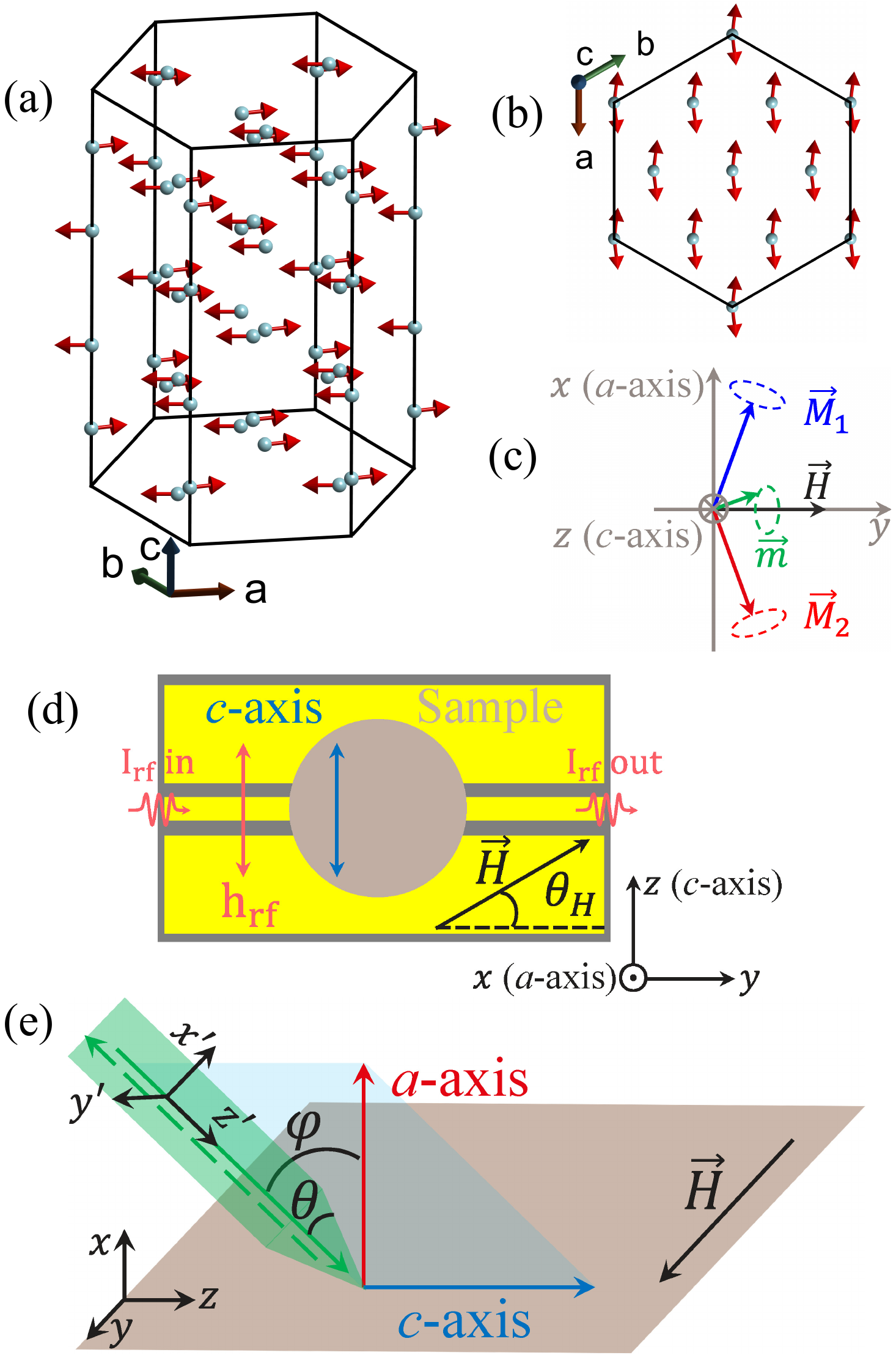}
	\caption{Hexagonal unit cell of the crystal structure of hematite from (a) the side and (b) top view. The cyan spheres and red arrows indicate the Fe atoms and the spins associated with them, respectively (oxygen atoms are not shown). We depict the canted antiferromagnetic state above the Morin temperature for which the sublattice spins lie in the $c$-plane along the $a$-axis with a small canting. They give rise to sublattice magnetization vectors $\mathbf{M}_1$ and $\mathbf{M}_2$. (c) Sketch of the low frequency quasi-ferromagnetic mode where a small magnetization (green arrow), $\mathbf{m}$, precesses elliptically around the applied field (black arrow), $\mathbf{H}$. (d) Schematics of the flip-chip VNA measurement. The sample (gray disk) is placed on a CPW. The static magnetic field, $\mathbf{H}$, is applied in the $ a $-plane and with an angle, $\theta_H$, to the  normal of the $ c $-axis of the crystal. The rf magnetic field (orange double-headed arrow), $\mathbf{h}_{\text{rf}}$, of the CPW is parallel to the $c$-axis. (e) Sketch of the BLS configuration. The magnetic field, $\mathbf{H}$, is applied perpendicular to $ c $-axis in the $ a $-plane. The laser light (green) forms a Gaussian beam and is focused on the surface of the sample (gray). The cone angle of the objective lens is $\theta$. The incident laser light with an incidence angle, $\varphi$ is scattered by magnons. We measure in  back-scattering geometry (dashed green arrow). }
	\label{fig1}
\end{figure}

\indent \textit{Introduction.}--- Spin waves (magnons) are collective spin excitations in magnetically ordered materials. They exhibit promising functionalities for information transmission and processing at GHz frequencies \cite{Kruglyak2010,Barman2021,Baumgaertl2022}. To realize energy efficient magnonic circuits \cite{Kruglyak2010,Barman2021,Tkachenko2012,Vogt2012,Mieszczak2020} isotropic spin wave (SW) dispersion relations, high group velocities, and low magnetic damping are essential. Until today, the artificial garnet Y$_3$Fe$_5$O$_{12}$ (YIG) \cite{Doughty1960} played a key role for the exploration of magnonics functionalities \cite{Serga2010}. Already in 1961, M. Sparks et al. coined the term that YIG was to ferromagnetic resonance research what the fruit fly was to genetics research \cite{Sparks1961}. This was particularly true for high-quality YIG grown by liquid phase epitaxy on the wafer scale \cite{Serga2010,Maendl2017}. However, in a ferrimagnetic material like YIG, magnon bands in the regime of small wave vectors, $k$, and low GHz frequencies are inherently anisotropic due to the dipolar interaction between spins. To overcome this, a lot of effort has been put into the development of microwave-to-magnon transducers which allow for the excitation of exchange dominated SWs with isotropic properties at high frequencies \cite{Yu2016,Baumgaertl2020,Che2020}. The transducers suffer however from typically a narrow bandwidth or require an applied magnetic field in contrast to conventional transmission lines and coplanar waveguides (CPWs).\\ \indent
In antiferromagnetic (AFM) materials, exchange interaction dominates the dispersion relation already at small wave vectors $k$. The dipolar interactions are virtually absent due to net zero magnetization. Still, SWs can propagate with high group velocities. Values similar to thick YIG \cite{Serga2010,Chumak2017} and as high as 30 km/s have been reported \cite{Turov1959,Keffer1966,Samuelsen1970,Hortensius2021}. However, the challenge with most AFMs is their net zero magnetization and sub-THz frequencies which make on-chip integration hard due to lack of efficient CPWs and THz sources (THz gap) \cite{Sirtori2002,Osborne2008}. Recently, the natural mineral and canted antiferromagnet hematite ($\alpha\text{-Fe}_2\text{O}_3$) \cite{Morrish1995} gained particular attention for magnonics \cite{Hamdi2022,Wang2022} after the observation of long distance spin transport \cite{Lebrun2018,Lebrun2020} and enhanced spin pumping \cite{Wang2021a}. It is known that, due to extremely low anisotropy in the basal plane \cite{Morrish1995,Pincus1960,Artman1965}, in the canted phase [Fig. \ref{fig1}(a) and (b)] one branch of the magnon modes resides at around 10 GHz at small $k$. 
Depending on the purity of hematite crystals, a damping coefficient as low as $7.8\times 10^{-6}$ was reported for the magnetic resonance \cite{Searle1968}. The hematite's finite net magnetization and strikingly small damping of below $10^{-5}$ make it hence suitable for magnonic applications. They allow for inductive coupling to CPWs and long-distance SW transport, respectively. However, there is no experiment reporting a measured SW dispersion for $k$ values accessible by CPWs with integrated microwave-to-magnon transducers  \cite{Yu2016,Baumgaertl2020,Che2020}. The dispersion measured over a large wave vector regime is of fundamental importance as it allows one to quantify the exchange stiffness length $l_e$ with large precision. $l_e$ is the key parameter to estimate the maximum possible spin-wave velocity of hematite in the GHz frequency regime.\\ \indent
Here, we study the magnon band structure of bulk hematite at different wave vectors $k$ by means of broadband microwave spectroscopy and $k$-resolved inelastic Brillouin light scattering (BLS) (Fig. \ref{fig1}). Using a CPW in flip-chip configuration we extract a magnetic damping parameter of $1.12\times 10^{-5}$ which is similar to the best YIG reported in Ref.~\onlinecite{Shone1985}. Still, the measured spectra with BLS show broad linewidths. Our modelling substantiates that the linewidth is explained by the SW dispersion relation, the large SW velocity and the Gaussian profile of the laser used for inelastic light scattering. The data substantiate a high group velocity of 10 km/s for $k=2.5$ rad/$\mu$m which are excited easily by a micron-sized CPW \cite{Wang2022,Watanabe2021}. In an applied field of 90 mT, the velocity increases to 16 km/s near $k=5$ rad/$\mu$m and levels off to 23.3 km/s for $k\geq 25$ rad/$\mu$m. The latter value has routinely been realized by transducers. Our findings substantiate hematite as a very promising candidate for a sustainable future of magnonics as its growth avoids the lead-based synthesis route used for high-quality YIG \cite{Doughty1960,Shone1985}.\\

\indent \textit{Properties of hematite.}--- We first briefly review the relevant magnetic properties of $\alpha\text{-Fe}_2\text{O}_3$. Hematite is the stable end product of oxidation of magnetite \cite{Moskowitz2015} and known for its great abundance as well as stability in an aqueous environment \cite{Iandolo2015}. It is an insulating antiferromagnet (AFM) with a corundum crystal structure. The arrangement of the magnetic atoms of Fe in the crystal is shown in Fig. \ref{fig1}(a) \cite{Hill2008}. At room temperature and above the Morin transition temperature of $T_\text{M}=262$ K the Fe$^{+3}$ magnetic moments lie in the $c$-plane due to an easy plane anisotropy, $H_{\text{A}}$, and stack antiferromagnetically along the $c$-axis [Fig. \ref{fig1}(a)] \cite{Artman1965,Morrish1995}. Within the $c$-plane, there is a weak 6-fold anisotropy around the $c$-axis, $H_{\text{a}}$, which favors the magnetic moments to align with the $a$-axes [Fig. \ref{fig1}(b)]. The magnetic moments of the two AFM sublattices are slightly canted away from the $a$-axis by the Dzyaloshinskii-Moriya (DM) interaction (Fig. \ref{fig1}(b)), resulting in a weak magnetic moment, $ \mathbf{m} $, perpendicular to both $a$- and $c$- axes at equilibrium \cite{Artman1965,Morrish1995}. The magnetization amounts to about 2 kA/m \cite{Moskowitz2015}. The magnetization dynamics of hematite in this weak ferromagnetic state offers two modes namely the quasi-ferromagnetic mode (qFM or low frequency mode) and quasi-antiferromagnetic mode (qAFM or high frequency mode) \cite{Pincus1960,Artman1965,Morrish1995}.
\\ \indent Here, we explore the qFM mode schematically depicted in Fig. \ref{fig1} (c). The canted AFM sublattice magnetization vectors precess elliptically around their equilibrium direction. This results in an elliptical precession of the weak magnetic moment, $ \mathbf{m} $ (green arrow in Fig. \ref{fig1} (c)), around the applied field, $\mathbf{H}$ \cite{Pincus1960,Artman1965,Morrish1995}. The frequency of the qFM mode was derived by Pincus \cite{Pincus1960,Morrish1995} according to
\begin{eqnarray}\label{Frequency}
	&&f_r=\\
&&\frac{\left| \gamma\right| \mu_0}{2\pi}\sqrt{H\sin\xi(H\sin\xi+H_{\text{D}})+2H_{\text{E}}(H_{\text{a}}+H_{\text{ME}})}, \nonumber
\end{eqnarray}
where, $\gamma$, $\mu_0$, $H_{\text{E}}$, $H_{\text{D}}$ and $H_{\text{ME}}$ is the electron gyromagnetic ratio, vacuum permeability,  exchange, DM and spontaneous magnetoelastic effective field, respectively. $\xi=\pi/2-\theta_H$ is the polar angle between $\mathbf{H}$ and the $c$-axis ($z$-direction). We define $\theta_H$ in Fig.~\ref{fig1}(d). Fink \cite{Fink1964} derived the dynamic susceptibility $\chi_{zz}(f,f_r)$ for the qFM mode. The real and imaginary parts read
\begin{eqnarray}\label{Susceptibility}
	\text{Re}\left[\chi_{zz}(f,f_r)\right]&=&\frac{\left(f_r^2-f^2\right)f_r^2}{\left(f_r^2-f^2\right)^2+\Delta f^2 f^2}~{\rm and}
	\\ \nonumber
	\\
	\text{Im}\left[\chi_{zz}(f,f_r)\right]&=&\frac{f f_r^2 \Delta f}{\left(f_r^2-f^2\right)^2+\Delta f^2 f^2},
\end{eqnarray}
respectively. Here, $f$ is the frequency of the radiofrequency (rf) magnetic field $\mathbf{h}_{\text{rf}}$ (Fig. \ref{fig1} (d)). The frequency linewidth, $\Delta f$, is related to the magnetic damping parameter, $\alpha$, by
\begin{eqnarray}\label{LineWidth}
	\Delta f\approx 2\alpha H_{\text{E}}(\gamma/2\pi),
\end{eqnarray}
for $H\ll H_{\text{E}}$. Unlike ferromagnets and uniaxial antiferromagnets, the resonance line width of a canted antiferromagnet does not depend on $f_r$ and is governed by the field-independent exchange frequency. Following Turov \cite{Turov1959,Keffer1966}, the SW dispersion of the qFM mode for hematite near $k=0$ is given by
\begin{eqnarray}\label{MagnonDispersion}
	&&f_m(k)=\\
	&&\frac{\left|\gamma\right|\mu_0}{2\pi}\sqrt{H(H+H_{\text{D}})+2H_{\text{E}}(H_{\text{a}}+H_{\text{ME}}+Ak^2)}, \nonumber
\end{eqnarray}
where $A=H_{\text{E}}l_{\text{e}}^2$ is the dispersion coefficient and $l_{\text{e}}$ is the effective magnetic lattice parameter or exchange stiffness length. For Eq. \ref{MagnonDispersion}, we considered the experimental geometry for $k$-resolved BLS measurements with an angle $\xi=\pi/2$ as described below.\\

\indent \textit{Experimental techniques.}--- The broadband microwave spectroscopy was conducted at room temperature above $T_\text{M}$ (Fig. \ref{fig1}(d)). The disk-shaped $a$-plane $\alpha\text{-Fe}_2\text{O}_3$ crystal had a diameter of 2.3 mm and thickness of 0.5 mm. Measurements were done in flip-chip configuration for which the sample was placed on a CPW with signal (ground) line width of 165 $\mu$m (295 $\mu$m). The $a$-axis of the crystal was perpendicular to the disk plane. The $c$-axis was in the plane and perpendicular to the CPW axis. Injecting a radio-frequency (rf) current, $I_{\text{rf}}^{\text{in}}$, into the CPW by port 1 of a vector network analyzer (VNA) induced the dynamic magnetic field $\mathbf{h}_{\text{rf}}$ (orange double-head arrow). The rf current was collected on the other end of the CPW by port 2 of the VNA ($I_{\text{rf}}^{\text{out}}$). The static magnetic field, $\mathbf{H}$, was applied in the $a$-plane of the crystal in all VNA measurements ($yz$-plane in Fig. \ref{fig1} (d)). For the angle dependent measurements an external field of 90 mT was applied and the angle, $\theta_{\text{H}}$, was varied in steps of $\Delta\theta_{\text{H}}=2^{\circ}$. In case of field sweep measurements, the applied field angle was perpendicular to the $c$-axis and in the $a$-plane of the crystal ($\theta_H=0~$deg). The field amplitude was varied in steps of $\Delta H=0.5$ mT.\\ \indent
Wave vector-resolved BLS measurements were done for $\theta_H=0~$deg on a piece of the same crystal in back-scattering geometry (Fig. \ref{fig1} (e)) using a green laser with wavelength, $\lambda=532~$nm and wave vector $k_0=2\pi/\lambda=11.81$ rad/$\mu$m. The external field was applied in the $a$-plane and perpendicular to $c$-axis. The sample for BLS was irregularly shaped. We ensured that it was tilted with respect to the incident laser beam along the $y$-axis in such a way that the laser beam remained in the $xz$-plane formed by the $a$- and $c$-axis. Since the penetration depth of the green laser in hematite is on the order of 75 nm \cite{Querry1984}, linear momentum conservation holds only for the in-plane component of the transferred wave vectors. Therefore, we define the transferred momentum from the light to magnons along the $c$-axis as $k_z=2(k_0\sin\varphi+k_{x}^{'}\cos\varphi)$, where $\varphi$ is the angle between the incident beam and the normal to the plane of the sample ($a$-plane) \cite{Sebastian2015}.
\begin{figure}[!ht]
	\centering
	\includegraphics[width=0.4\textwidth]{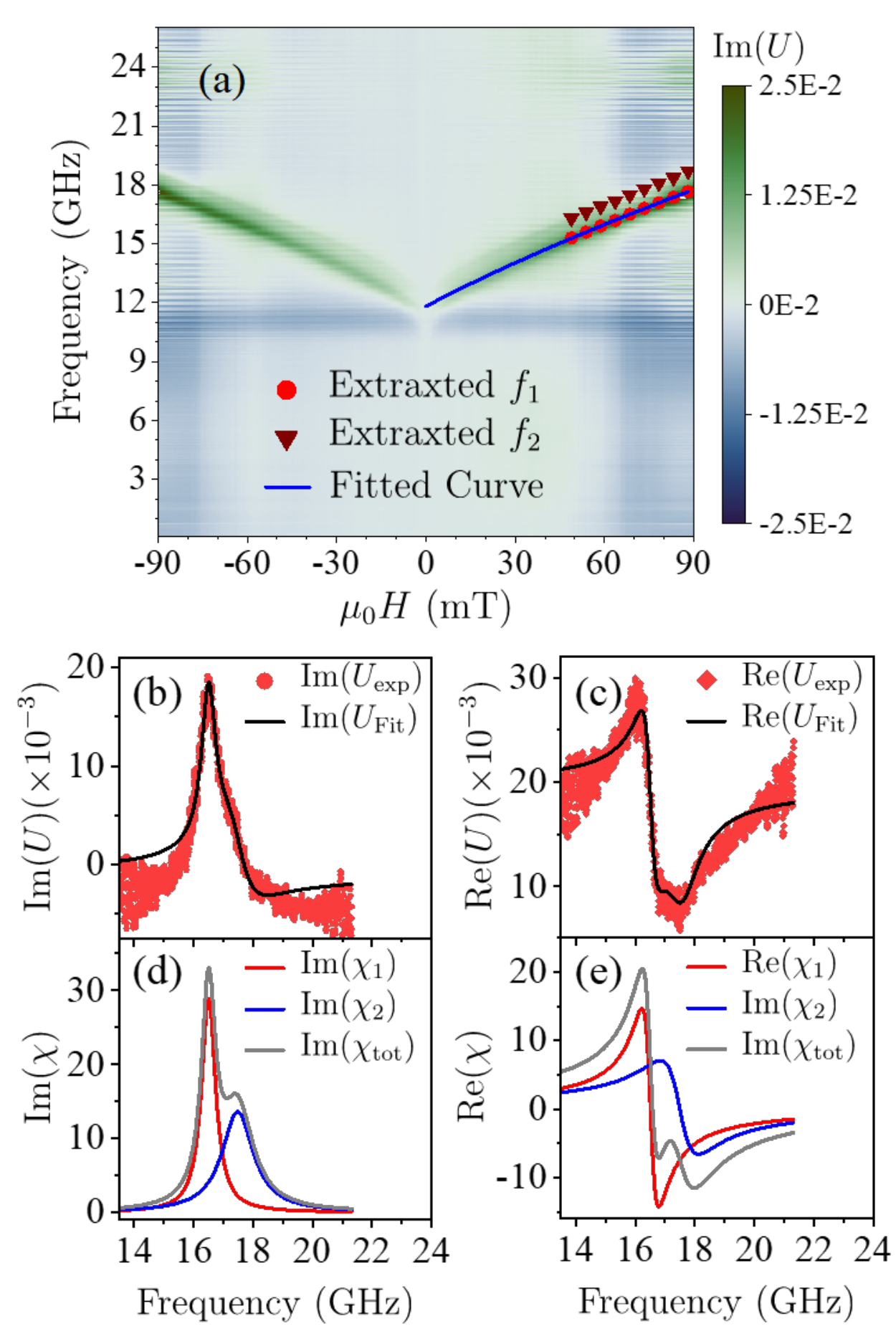}
	\caption{(a) Color-coded field dependent $\text{Im}(U)$ parameter measured on the sample as indicated in Fig. \ref{fig1}(d) with $\theta_H=0$~deg. Solid red circles and wine triangles are the frequencies extracted by fitting Eq. \ref{Ufitt} on the data. Blue solid line is obtained by fitting Eq. \ref{Frequency} with extracted $f_1$ values. (b) Imaginary and (c) real part of $U_{\text{exp}}$ (red symbols) and $U_{\text{fit}}$ (black curves) for 70 mT, respectively. (d) Imaginary and (e) real parts of $\chi_1$ (red curve), $\chi_2$ (blue curve) and their sum (gray curve) for 70 mT, respectively.}
	\label{fig2}
\end{figure}
We assume a Gaussian beam profile giving rise to a Fourier transform of the beam intensity $I$ as
\begin{eqnarray}\label{Gauss}
	I(k_{x}^{'})= e^{k_{x}^{'2} w_{0}^{2}/2}
\end{eqnarray}
with $w_0=\frac{2}{k_0 \text{NA}}$. $NA$ is the numerical aperture of the lens \cite{Novotny2012,Levy2019}. The momentum $k_{x}^{'}$ has a projection on $z$-direction due to the focusing of the beam.\\

\begin{figure}[!ht]
	\centering
	\includegraphics[width=0.37\textwidth]{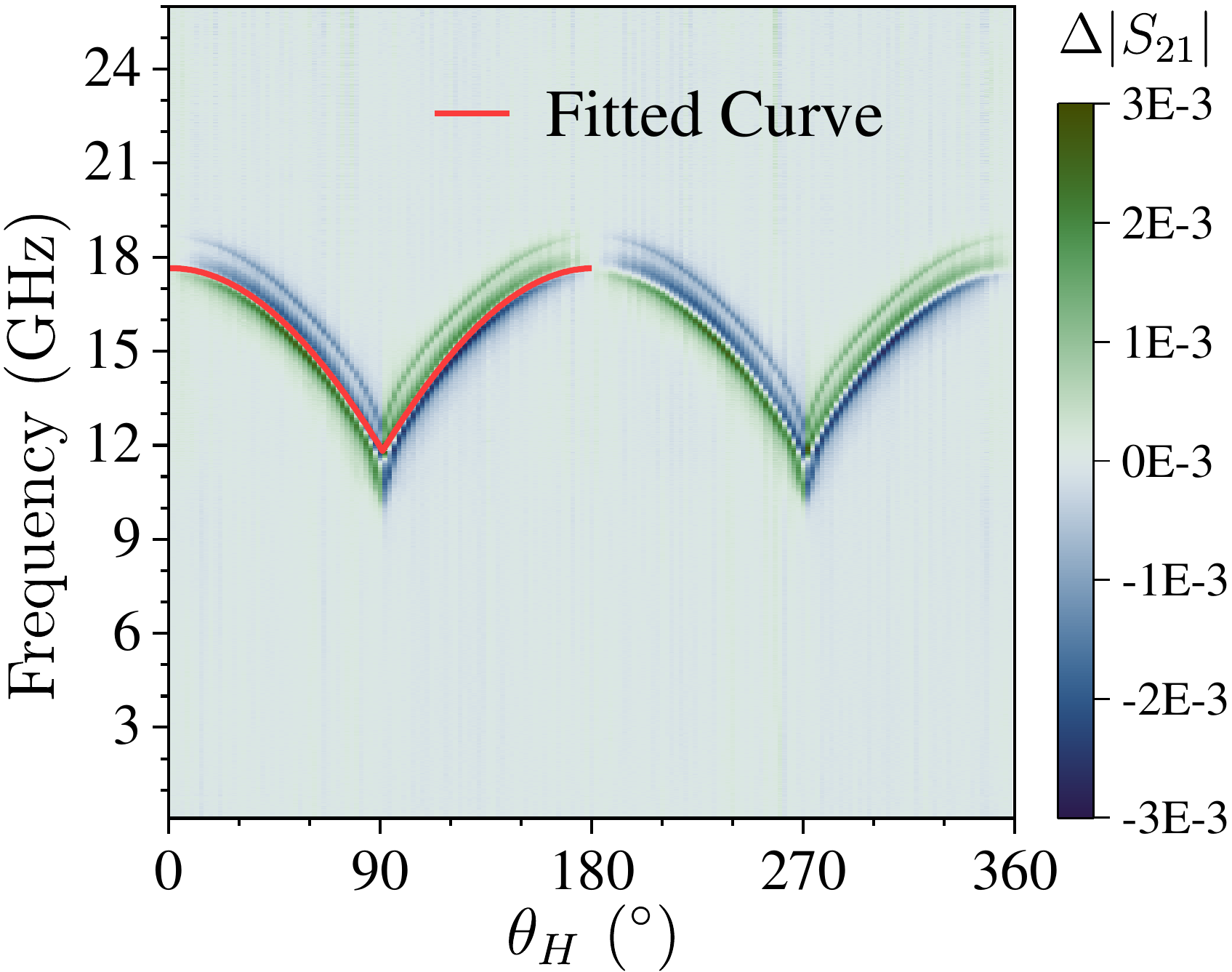}
	\caption{Color-coded neighbor subtracted angle dependent VNA-FMR spectra measured on the sample as indicated in Fig. \ref{fig1}(d) with $\mu_0H=90$ mT. Orange curve depicts Eq. \ref{Frequency} by extracted material parameters.}
	\label{fig3}
\end{figure}

\indent \textit{Broadband microwave spectroscopy data.}--- Field-dependent VNA spectra are shown in Fig. \ref{fig2} (a). We depict the imaginary part of the quantity $U(f,H)=i\ln\left[S_{21}(f,H)/S_{21}(f,H=0) \right]$ in a color-coded plot, where $S_{21}$ is the complex scattering parameter measured by the VNA at a given field, $H$. We identify two branches which we label $f_1$ and $f_2$ for positive fields. For a detailed analysis, we consider that the parameter $U$ contains the susceptibility $\chi$ of the sample \cite{Kalarickal2006,Bilzer2007} and the electromagnetic response of the rf circuit used in the flip-chip method. To account for the different contributions, we follow Refs.  \cite{Kalarickal2006,Bilzer2007} and fit the measured $U$ with
 \begin{eqnarray}\label{Ufitt}
	&&U_{\text{fit}}=\\
	&&C[1+\chi_{0}+\chi(f,f_1,\Delta f_1)e^{i\phi_1}+\chi(f,f_2,\Delta f_2)e^{i\phi_2}] \nonumber
\end{eqnarray}
as shown in Fig. \ref{fig2}(b) and (c) (black lines). $C$ is a real-numbered scaling parameter, $\chi_{0}$ is a complex-numbered offset parameter, and $\phi_i$ are phase shift adjustments ($i=1,2$). Using Eq.~\ref{Ufitt}, we extract the resonance frequencies and linewidths $\Delta f_i$ for the two modes labelled by $f_1$ and $f_2$. In Fig. \ref{fig2} (b) and (c), we display the measured imaginary and real parts of $U$ with red symbols. In Fig. \ref{fig2} (d) and (e), we show the extracted real and imaginary parts together with the total susceptibilities (gray lines) from which the two resonant modes are identified. Their frequencies extracted for different $H$ are depicted in Fig. \ref{fig2} (a) with solid red circles and magenta triangles. We focus on fields larger than 50 mT to ensure a saturated state.\\ \indent
The blue line in Fig. \ref{fig2}(a) results from fitting Eq. \ref{Frequency} to branch $f_1$ with $\xi=\pi/2$. From the fit, we obtain $\mu_0H_{\text{E}}=1003.61$~T, $\mu_0H_{\text{D}}=2.34$~T and $\mu_0(H_{\text{a}}+H_{\text{ME}})=88.64~\mu$T. Using Eq. \ref{LineWidth} and the experimental value of $\Delta f_1=0.63$ GHz for branch $f_1$, we extract a damping parameter of $\alpha=1.12\times 10^{-5}$. This value is similar to the best value reported for YIG in Ref.~\cite{Shone1985}. The branch $f_2$ will be discussed later.\\
\indent Angle-dependent VNA spectra are shown in Fig. \ref{fig3}. 
To enhance the signal to noise ratio, we depict $\Delta \left|S_{12} \right| = \left|S_{12}(H+\Delta H)- S_{12}(H) \right|$. For such data, a zero-crossing (highlighted by the red curve) represents the resonance frequency. The spectra show a 2-fold symmetry as expected for the $c$-axis being in the plane of the applied magnetic fields. Introducing the extracted parameters discussed above, Eq. \ref{Frequency} models well the resonance frequency of branch $f_1$ as a function of angle $\theta_H$ (red line). The angular dependency is hence consistent with the effective anisotropy field extracted from the field-dependent data.\\

\begin{figure}[!t]
	\centering
	\includegraphics[width=0.45\textwidth]{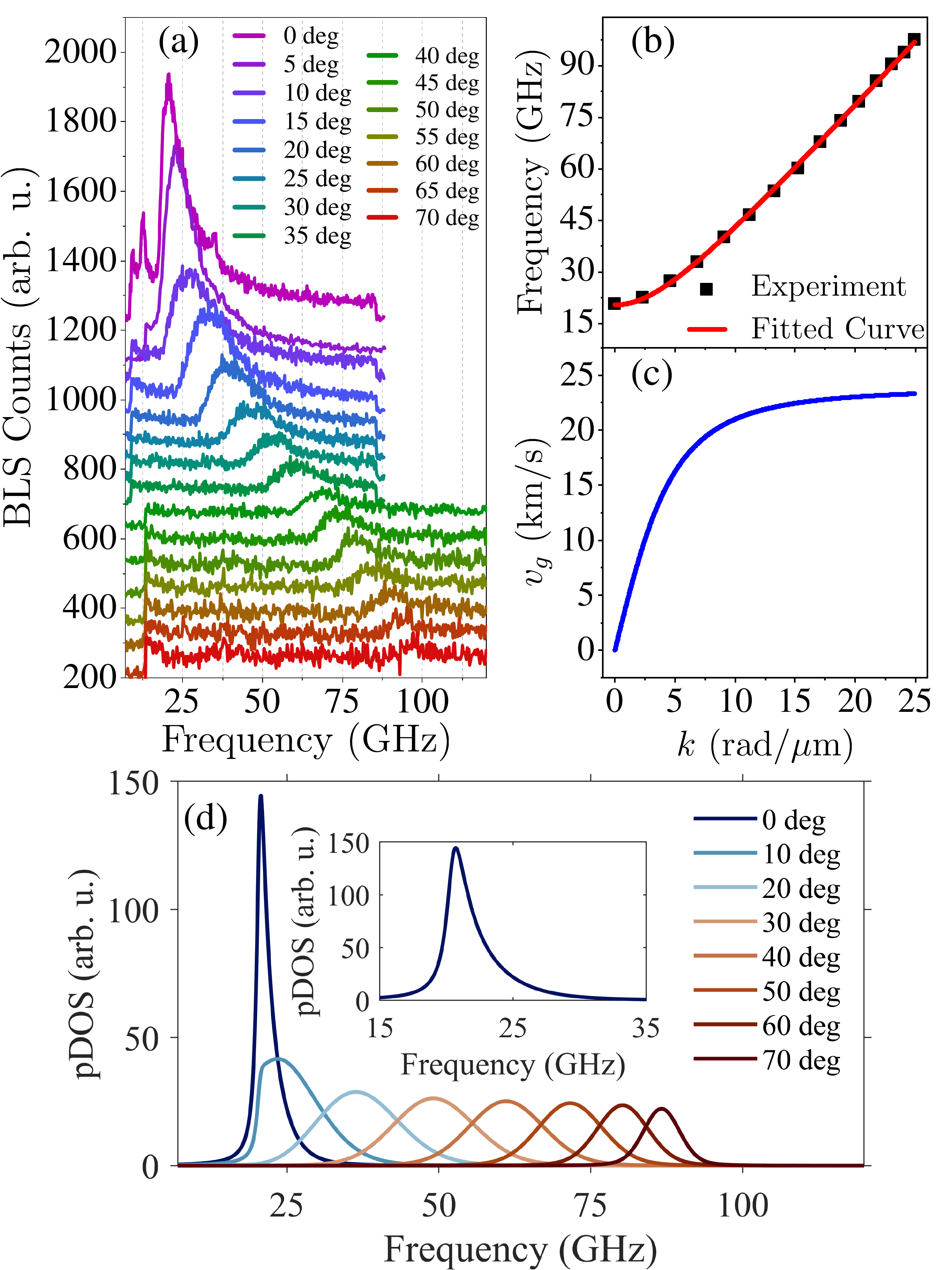}
	\caption{(a) BLS spectra for different incident angles, $\varphi$, measured as indicated in Fig. \ref{fig1}(e) with $\mu_0H=90$ mT. (b) Extracted peak maxima frequency (black squares) as a function of corresponding transferred wave vector, $k$ and fitted dispersion relation (red curve) by Eq. \ref{MagnonDispersion}. (c) Group velocity (d) partial density of states and obtained from fitted magnon dispersion at $\mu_0H=90$ mT. Inset in (d) depicts pDOS for $\varphi=0$ degree.}
	\label{fig4}
\end{figure}

\indent \textit{BLS data.}--- The BLS spectra for different values of incident angle, $\varphi$, are shown in Fig. \ref{fig4} (a). Black symbols in Fig. \ref{fig4} (b) depict the frequency position of the maximum BLS peak as a function of wave vector calculated from the incidence angles shown in the legend of Fig. \ref{fig4} (a). The red curve in Fig. \ref{fig4} (b) is obtained by considering first the material parameters obtained from VNA measurements and then fitting Eq. \ref{MagnonDispersion} to the black symbols in the same graph. We obtain a dispersion coefficient of $\mu_0 A=9.153\times 10^{-6}$ T.$\mu$m$^2$ which leads to an exchange stiffness length of $l_{\text{e}}=0.955$ \AA. Using Eq. \ref{MagnonDispersion} and the obtained parameters we calculate the SW group velocity, $v_{\rm g}$, for $\mu_0 H=90~$mT [blue line in Fig. \ref{fig4} (c)]. The velocity $v_{\rm g}$ increases significantly with $k$ and levels off at 23.3 km/s. Such a high group velocity is the direct result of the strong exchange interaction and at the same time vanishingly small net magnetization. Microstructured CPWs used in magnonics offer spin-wave wave vectors around $k=2.5$ rad/$\mu$m \cite{Wang2022,Watanabe2021}. For such a value $k$, the qFM spin wave in hematite exhibits a group velocity of $v_{\rm g}=10$ km/s similar to SWs in thick YIG and about a factor of 10 larger compared to ultra-thin YIG \cite{yu2014magnetic}. The decay length of the SWs is given by $l_{\rm d}=v_{\rm g} \tau$, where $\tau=(1/2\pi\alpha f_{m}(k))$ is the relaxation time of the SW. For 30 mT applied field and $k=2.5$ rad/$\mu$m we calculate $v_{\rm g}=13.4$ km/s and $\tau=840.6$ ns which leads to $l_{\rm d}=11.3$~mm, again similar to thick YIG. This is due to low damping and high group velocity of the qFM spin waves in hematite.
\\ \indent
Despite the small damping of the hematite sample measured by the VNA, the resonance peaks of the qFM mode in the BLS spectra show a large width of 10 to 20 GHz. Furthermore, the BLS peak taken at nearly normal incidence of the laser beam (e.g. at $\varphi=0$ or $5$ deg) shows a strong asymmetric shape. As $\varphi$ increases, the intensity of the prominent peak reduces and it becomes more and more symmetric. To understand these observations, we calculated the partial density of states (pDOS) by considering the magnon dispersion relation and the Gaussian beam profile according to
\begin{eqnarray}\label{pDOS}
	&&pDOS(f,\varphi)=
	\\
	&&-\frac{1}{\pi}\int_{-\infty}^{+\infty}d{k_{x}^{'}}\, \cos\varphi\, \text{Im}\left[\frac{I(k_{x}^{'})}{f-f_m(k)+i\Delta f}\right]. \nonumber
\end{eqnarray}
Assuming only fully back reflected light we have $k=2(k_0\sin\varphi+k_{x}^{'}\cos\varphi)$. $\Delta f$ is the resonance broadening due to the damping given by Eq. \ref{LineWidth}. The numerical aperture of the objective lens was $NA=0.18$. The calculated pDOS for different incident angles is plotted in Fig. \ref{fig4} (d). Comparing Figs. \ref{fig4}(a) and (c), the experimental BLS peaks are broad because they contain a certain frequency regime of the band structure which is determined by the Gaussian wave vector distribution of $k=2k_0\sin\varphi+\Delta k$. The central wave vector is $k_c=2k_0\sin\varphi$ and the distribution function for $\Delta k=2k_{x}^{'}\cos\varphi$ is given by $I(k_{x}^{'})$ (Eq. \ref{Gauss}). We attribute the broad BLS peaks hence to the high group velocity of magnons which leads to a peak width proportional to $v_g\times \Delta k$. The asymmetry of BLS peaks for small $\varphi$ can be understood in terms of the Van Hove singularity of the pDOS near $k=0$. As $\varphi$ increases, the part of the band structure which is relevant for the collected light shifts away from $k=0$. Consequently, the peak gets more symmetric with $\varphi$. The peak intensity reduces with $\varphi$ due to the factor $\cos\varphi$ in Eq. \ref{pDOS}. We note that Eq. \ref{pDOS} does not include all possible scattering processes. Still it provides a good qualitative understanding about the contributions giving rise to the characteristics shape and signal strength of BLS peaks as a function of $\varphi$.\\

\indent \textit{Discussion.}--- The branch of mode $f_2$ in Fig. \ref{fig2}(a) is higher by about $\delta f=1$~GHz than $f_1$ at the same field $H$. A second branch was reported also in Ref. \onlinecite{Bialek2022}. Here the authors studied the qAFM mode at zero field and assumed a distribution of magnetic domains. We applied a large enough magnetic field to avoid domains. A domain formation can not explain our second branch. Another possibility for $f_2$ is a standing wave along the thickness of the sample or a SW excited with a discrete wave vector coming from the CPW. However, for these cases a quantitative estimate based on the magnon dispersion obtained with BLS (Eq. \ref{MagnonDispersion}) led to a frequency separation of much smaller than 1 GHz. \\
\indent The remaining explanation for $f_2$ is a nonuniform magnetoelastic field in the sample induced when fixing the sample on the CPW. We attribute the observed frequency splitting to different strains in different parts of the sample that is either in contact with the CPW conductors or floating on the CPW gaps. A difference of $\delta H_{\text{ME}}=24~\mu$T would account for $\delta f=f_2 - f_1=1$~GHz. The effect of magnetoelastic interaction by unidirectional compression, $\mathbf{p}$, in the basal plane of the crystal, on the qFM mode is given by replacing $H_{\text{ME}}$ with $H_{\text{ME}}^{\prime}=H_{\text{ME}}-Rp\cos2\psi$ \cite{Levitin1969,Maksimenkov1974}. Here, $R=287$ $\mu$T/bar \cite{Maksimenkov1974} is a coefficient determined by the elastic and magnetoelastic parameters of the crystal and $\psi=\pi/2$ is the angle between $\mathbf{p}$ and $\mathbf{H}$. Using $\delta H_{\text{ME}}=Rp$ we obtain $p=0.084$ bar. This value corresponds to a force of 14.6 mN on the parts of the crystal that are in contact with the CPW conductors. The evaluated force is one order of magnitude smaller than the weight of our sample and indicates the known sensitivity of hematite towards magnetoelastic effects. \\
\indent From our experiments, we do not determine $H_{\text{a}}$ and $H_{\text{ME}}$ separately as they enter Eq. \ref{Frequency} in the same way. To separate the small $H_{\text{a}}$ from $H_{\text{ME}}$ an angle dependent VNA measurement with the magnetic field applied in the $c$-plane of the crystal would be required. The existing sample was not suitable for that.\\
\indent Our BLS data were acquired on a seven times larger wave vector regime compared to the data used in Ref.~\onlinecite{Wang2022} where, the authors extracted an exchange stiffness length $l_{\text{e}}$ of $1.2~$\AA. From our extended dataset, we evaluate the value $l_{\text{e}}=0.96$~\AA. The precise determination of $l_{\text{e}}$ is of key importance as it determines the SW group velocity. Contrary to the ferrimagnetic YIG, dipolar effects are not expected to play an important role in SW dispersion of hematite. Hence, hematite thin films can provide similarly large spin-wave velocities as reported here for the bulk crystal. As a consequence, they potentially outperform YIG thin films concerning speed and decay lengths at wave vectors which are realized by the state-of-the-art transducers. Furthermore, hematite is based on earth abundant elements and as an end product of oxidation of magnetite a stable natural mineral suggesting sustainable synthesis routes.\\

\indent \textit{Conclusion.}--- We have measured the magnon dispersion relation of hematite for wave vectors $k$ which are relevant for timely experiments in magnonics. The damping coefficient of the studied natural crystal was $1.1\times 10^{-5}$ at room temperature. This value is only 40~\% larger than the best value reported for pure hematite and is already as good as the best YIG. The estimated spin-wave decay length for $k=2.5~$rad/$\mu$m is larger than 1 cm in a small magnetic field. In optimized thin films, current microwave-to-magnon transducers are expected to achieve larger group velocities in hematite than in YIG. The reported properties suggest that hematite can become the fruit fly of sustainable modern magnonics.\\

\indent \textit{Note added.}--- While completing the manuscript about our experiments on hematite \cite{Hamdi2022}, we became aware of Ref.~\onlinecite{Wang2022}. The authors measured group velocities and the spin-wave dispersion via integrated CPWs for 1~rad$/\mu$m$~\le k\le3.5~$rad$/\mu$m. Our BLS measurements cover a seven times larger regime of $k$ values enabling an improved evaluation of the parameter $l_e$. This parameter is decisive to estimate the saturation velocity of GHz SWs in hematite.\\ \indent The scientific colour maps developed by Crameri et. al. \cite{Crameri2021} is used in this study to prevent visual distortion of the data and exclusion of readers with colour-vision deficiencies \cite{Crameri2020}.\\

\indent \textit{Acknowledgement.}--- The authors thank SNSF for financial support via grant 177550. We acknowledge discussions with A. Mucchietto, M. Mruczkiewicz, S. Nikitov and A. Sadovnikov. The different pieces of the hematite crystal were provided by J.-P. Ansermet and M. Bialek. We thank them for the support and discussions.


%

\end{document}